\documentstyle[preprint,aps,epsf,floats,tighten]{revtex}
\input{epsf.sty}
\begin{document}
\draft
\title{A systematic analysis of new physics effects in spin correlations
in $q\overline q\rightarrow t\overline t$}
\author{Tibor Torma\thanks{kakukk@physics.utoronto.ca}}
\address{Department of Physics, University of Toronto, 
Toronto, Ontario, Canada M5S~1A7}
\date{\today}
\maketitle
\begin{abstract}
I show that there are eight new physics operators that contribute to top
pair production and subsequent decay in interference with the leading SM process at
the Tevatron and estimate their expected statistical inaccuracies. All fully
differential cross sections are calculated in the \mbox{6-particle} phase space. The
observation of angular correlations dramatically improves the accuracy of the
extraction.
\end{abstract}
\narrowtext

\section{Introduction}

Until high energy experiments achieve the neccesary energy and sensitivity to
discover new particles in direct production, any experimental progress beyond the
Standard Model (SM) should go through the observation of loop effects in precision
measurements. These new effects can be parametrized using effective Lagrangians
where coefficients of higher dimensional operators carry all the information on
new physics. In this paper I ask how accurately one can measure these couplings in
proton-proton collisions at presently available energies.

The heavy mass of the top quark allows us to speculate that new physics could show
up more easily in the couplings of the top quark then of any other SM particle. The
most obvious place do such precision measurements is the Fermilab Tevatron, where at
Run~2 an estimated 20,000 top pairs will be produced. In this paper I am looking
for small new operators, generated in many theories with strong electroweak
physics, large extra dimensions, etc.  The succes of the SM leads us to expect
that these contributions are small, and this in turn allows us to keep only the
interference with the leading SM process, top pair production through $q\overline
q\rightarrow g\rightarrow t\overline t$. The emerging restriction singles out eight
different operators and these can be disentengled once we make use of the
information  residing in the spin state of the produced top quarks. This spin
information is in turn translated into the angular distribution of the decay
products of the two $t,\overline t$ quarks.

The effect of new physics operators in top pair production has been repeatedly
studied in the literature~\cite{Rizzo,Hikasa,HillParke,Peccei2,Cheung}. Most of
these investigations look at top transverse momentum and invariant $t\overline t$
distributions in addition to effects in the total cross section. In the following
I argue that this approach looses most of the information present in the events
because it sums over top polarization states. The top quark decays into
$t\rightarrow Wb$ so quickly that its spin state is not washed out by strong
interactions. In Refs.~\cite{Hikasa,Cheung} the use of particular asymmetries helps
to incorporate at least some of this information. We will see however, that the full
use of angular distributions significantly increases the accuracy.

The first attempt in this direction was taken in~\cite{HoldomTorma}, where the effect
of the top chromomagnetic moment on these distributions was analyzed. We now extend
this analysis for all operators that can contribute, up to dimension~6. In fact,
there is little reason, other than theoretical prejudice based on adherence to
particular sorts of models, to expect that any of the operators in the set should be
distinctly larger than others. For example, the dimension~5 chromomagnetic
moment~$\kappa$ is produced at ${\cal O}(\frac{1}{\Lambda^2})$, similarly to
dimension~6 operators ($\Lambda$ is the scale of the new physics.) In this paper I
purposefully avoid any reference to the expected relative size of the new couplings.

In what follows I classify the observable contributions in two cases: (i) when
the electroweak (EW) symmetry breaking sector contains only a light Higgs doublet,
and (ii)~when all new fields are heavy. In the former case we can
write an effective Lagrangian where the fields are in a linear representation of the
EW group, while in the latter case we must use a nonlinear representation. The
approach is thus quite general and should only break down if there are new
resonances in the below-$1\,TeV$ region that interact significantly with the top
quark (such as top pions, for example.) We write down the new operators in both
cases and conclude that this experiment cannot tell apart the two scenarios.

The paper is organized as follows. In Sec.~\ref{sec:reprezes} I first discuss the
relationship of the two representations, and write down a basis of those new
operators that contain at least one top field. Then I find the corrections to the
$gt\overline t$ vertex, to top decay in the $tWb$ vertex and 4-quark vertices: two
CP violating and six CP conserving operators. I proceed in Sec.~\ref{sec:diffsigma}
by calculating the contribution of each of these new interactions to the $t\overline
t$ spin density matrices and to top decay, and compute their contribution to the
fully differential cross section in the 6-particle phase space. These contributions
can be used for comparing experimental distributions to disentangle each
contribution.

It is important to note that one does not need to reconstruct the probability of
events in the many-dimensional phase space from the measurement. That would be
impossible with the event numbers involved (on the order of several thousand). One
can instead estimate the new couplings, using for example a Maximum Likelihood
method. We perform such an analysis in~Sec.~\ref{sec:sens}, using
Monte-Carlo-generated data according to the SM distribution. This analysis tells us
the attainable statistical accuracy of the estimation.

I sum the results in the Conclusion and discuss the discovery potential of
TeVatron Run~2 at Fermilab. I find that new physics at a scale order
$TeV$ can be generically detected and that all-hadronic decay modes of the top
quarks may play an important role, the complexity of these events notwithstanding.

\section{Linear and nonlinear representations: parametrization of new physics in
interference}\label{sec:reprezes}

We are interested in effective Lagrangians controlling top pair production and
subsequent decays with or without a single SM Higgs field, where all
other physics resides in the higher dimensional operators. Without including any
model dependent ``prejudice", we should proceed by writing down all possible
operators at each dimension level and look for their experimental consequences.

When a SM Higgs boson is present in the Lagrangian, the underlying
$SU(2)\times U(1)$ symmetry imposes restrictions on the possible form of the new
operators. We are free to choose the fields in a linear representation of the gauge
group and implement spontaneous electroweak breaking by introducing a Higgs vacuum
expectation value. The introduction of a Higss v.e.v however induces lower
dimensional new operators of a non-gauge invariant form.

It may well happen that the SM Higgs is very heavy or is not present at
all. In the absence of light resonances an effective Lagrangian still make sense,
but in that case its field content does not allow us to put all the fields in closed
linear representations. Because we are supposing that all additional fields are
heavy, we are allowed to put the SM fields, including the three eaten-up
Goldstone bosons, in a nonlinear representation of the gauge group and require gauge
invariance of the effective Lagrangian.~\cite{Peccei1}

It has been shown~\cite{Burgess}, however, that the additional restrictions
by imposing this nonlinearly realized symmetry are exactly compensated by the
additional freedom in the choice of the gauge. Note that the presence of unphysical
degrees of freedom in a generic gauge (such as longitudinal and timelike components
of the gauge bosons) lets us write down additional operators containing these
fields. In the unitary gauge all these nonphysical fields fall away and the net
result is that all restrictions on the new operators from gauge invariance disappear.
The gauge non-invariant operators allowed this way will be similar to those in the
linear representation that contain no Higgs fields.

In this section I write down, in both cases, all the new physics operators that can
effect the $q\overline q\rightarrow t\overline t$ process, up to dimension~6. There
is a huge number of such operators. However, one may impose the following
restrictions on those that may give appreciable contributions:
\begin{itemize}
\item{The use of an effective Lagrangian presupposes
that the corrections are small, therefore we are justified to keep only the
interference between new physics operators and SM processes.}
\item{We keep new physics only in vertices that involve at least one top quark
field. Other new physics, including that in light quark interactions, can presumably
be seen easier in other processes.}
\item{We drop those new operators that pick up a
weak coupling from an unmodified SM vertex (such as corrections to the
$Zt\overline t$ vertex, contributing to $q\overline q\rightarrow Z\rightarrow
t\overline t$).}
\item{We drop all loop corrections.}
\end{itemize}

\begin{figure}
\begin{center}
\mbox{\epsfxsize=6in \epsfbox{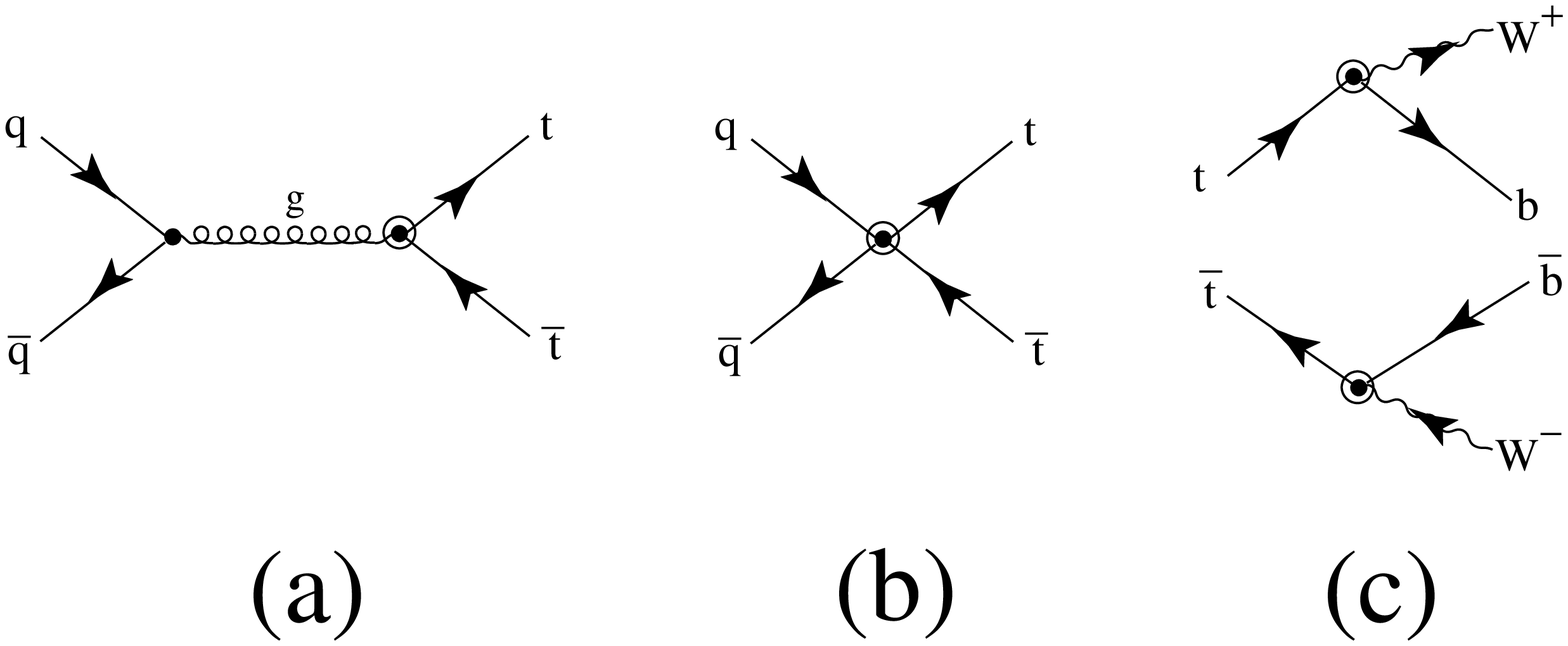}}
\end{center}
\caption{\protect\label{fig:remains}The only types of new top physics processes that
contribute to $q\overline q\rightarrow t\overline t$ in interference.}
\end{figure}

With the above criteria some combinatorics shows that the only contributions come
from (a) corrections to the $gt\overline t$ vertex, (b) chirality preserving 4-quark
operators of the color octet type,
$\left(\overline q\,\Gamma_{V,A}\frac{\lambda^a}{2}q\right)\left(\overline
t\,\Gamma_{V,A}\frac{\lambda^a}{2}t\right)$ where the light quarks  $q,\overline q$
are must be of the same flavor  to interfere with the SM process, and
(c) corrections to the $tWb$ vertex in top decay (see Fig.~\ref{fig:remains}).

The linear set of 2-quark (plus boson fields) operators have been
classified extensively in the literature~\cite{Whisnant,Hikasa,more lists}. Any
complete prediction of the contributions makes sense with reference to a complete
set of operators, because any partial list will tacitly imply a particular choice of
operators that have zero contribution. For this reason I provide in
Appendix~\ref{app:lin} a list of the operator basis I am using. This list
completely agrees with that of Ref.~\cite{Whisnant} when all CP-odd operators
are dropped. In Appendix~\ref{app:nonlin} I also provide the corresponding
operators in the nonlinear set.

The contributions we find to the $gt\overline t$ vertex are only:
\begin{eqnarray}\label{eq:gtt}
-g_s\frac{\kappa}{4m_t}\overline
t\,\sigma^{\mu\nu}\,\frac{\lambda^a}{2}G^a_{\mu\nu}\,t
-g_s\frac{\tilde\kappa}{4m_t}\overline
t\,\sigma^{\mu\nu}\,\frac{\lambda^a}{2}{\rule{0pt}{1.7ex}^*}G^a_{\mu\nu}\,t,
\end{eqnarray}
where the nonlinear set contributes $\kappa=\kappa_g^{33}$ and
$\tilde\kappa=\tilde\kappa_g^{33}$, while the linear set contributes
$\kappa=-4m_t\,\frac{v}{\sqrt2}\,\mbox{Re}\,b_{10}^{33}$ and
$\tilde\kappa=4m_t\,\frac{v}{\sqrt2}\,\mbox{Im}\,b_{10}^{33}$ 
(for the detailed meaning of the r.h.s. see the Appendix.) Both $\kappa$ and
$\tilde\kappa$ are required to be real by Hermiticity. The absence of the ${\cal
O}(p^2)$ terms in the expansion of the $gt\overline t$ form factors seems rather
strange. These operators have been turned into 4-quark operators by the use of the
equations of motion.

Next we look at corrections to top decay. We find the following three operators
that seem to contribute to the $tWb$ vertex:
\begin{equation}\label{eq:tWb}
\begin{array}{ccl}
\frac{c_1}{m_t}\,\frac{g}{\sqrt2}\,W^\mu\,\overline
t_R\,i\stackrel{\leftrightarrow}{\partial}_\mu b_L+h.c. 
 &\hspace{3em}&c_1=\left\{
\begin{array}{cl}
-\frac{\kappa_{LW}^{\prime33}}{4}&\mbox{(nonlinear)}\\&
\\\frac{m_t}{2}\frac{v}{\sqrt2}\,\overline{b_2^{33}}&\mbox{(linear)}
\end{array}
\right.\\&&\\
-\frac{c_2}{m_t}\,\frac{g}{\sqrt2}\,W_{\mu\nu}\,\overline
t_R\,\sigma^{\mu\nu} b_L+h.c. 
 &&c_2=\left\{
\begin{array}{cl}
\frac{\kappa_{W}^{33}+i\,\tilde\kappa_{W}^{33}}{4}&\mbox{(nonlinear)}\\&
\\-m_t\,\frac{v}{\sqrt2}\,\overline{b_9^{33}}&\mbox{(linear)}
\end{array}
\right.\\&&\\
i\frac{c_3}{m_t^2}\,\frac{g}{\sqrt2}\,\left(\partial^\mu W^\nu+\partial^\nu
W^\mu\right)\,\overline t_L\,\gamma_\mu\,i\stackrel{\leftrightarrow}{\partial}_\nu
b_L+h.c. 
 &&c_3=\left\{
\begin{array}{cl}
-i\rho_{LW}&\mbox{(nonlinear)}\\&
\\0&\mbox{(linear)}
\end{array}
\right.
\end{array}
\end{equation}
In the following we will see however, that only three real linear combinations of
these operators contribute to the differential cross section; the operator with
$c_3$ does not contribute at all. It is not unexpected to see fewer operators
contribute to top decay because of the kinematical restrictions. Only the CP-even
combination
\begin{equation}
\mu_1=2\frac{m_t^2-m_W^2}{m_t^2}\,\mbox{Re}\left(2c_2+c_1+c_3\right)
\end{equation}
and the CP-odd combination
\begin{equation}
\mu_2=2\frac{m_t^2-m_W^2}{m_t^2}\,\mbox{Im}\left(2c_2+c_1+c_3\right)
\end{equation}
affect the differential cross sections, while the CP-even
\begin{equation}
\mu_{0}=-2\frac{m_t^2-m_W^2}{m_t^2}\,\mbox{Im}\left(c_1+c_3\right)
\end{equation}
only changes the total rate by an overall factor of $(1+\mu_0)$. Because this
last factor would show up only in the total top decay rate, it is irrelevant for
our investigation and I drop it in the following. Note that in a similar manner I
ignored the contribution of a  \(\sim W^\mu\,\overline t_L\,\gamma_\mu\,b_L\)
operator which would only modify~$V_{tb}$.

Finally, we turn to the effect of 4-quark operators. These are the same in both set
and are severely restricted by the above criteria:
\begin{eqnarray}\label{eq:4F}
\frac{g_s^2}{4m_t^2}&\times&\left\{\,
c_{VV}\times \left(\overline q\gamma_\mu\frac{\lambda^a}{2} q\right)\,\left(\overline
  t\gamma^\mu\frac{\lambda^a}{2}t\right)+
c_{VA}\times \left(\overline q\gamma_\mu\frac{\lambda^a}{2} q\right)\,\left(\overline
  t\gamma^\mu\gamma^5\frac{\lambda^a}{2}t\right)\right.
\\&&\left.\ +
c_{AV}\times \left(\overline q\gamma_\mu\gamma^5\frac{\lambda^a}{2} q\right)\,
  \left(\overline t\gamma^\mu\frac{\lambda^a}{2}t\right)+
c_{AA}\times \left(\overline q\gamma_\mu\gamma^5\frac{\lambda^a}{2}q\right)\,
\left(\overline t\gamma^\mu\gamma^5\frac{\lambda^a}{2}t\right)
\,\right\},\nonumber
\end{eqnarray}
where the coefficients $c_{VV},c_{VA},c_{AV},c_{AA},$ must all be real
($q=u,d,s,...$ is the light quark). The selection of these operators has been
pressed on us by the required presence of interference with the SM process. However,
our operators can be embedded into the set of custodial invariant 4-fermion
operators discussed in~\cite{HillParke}. Therefore, we find no direct restriction on
our operators from custodial symmetry. I separated off a factor of $g_s^2$ for easy
comparison (this factor does not need to reflect any strong interaction physics.)

This completes the list of all contributing operators.
Their size remains remarkably unrestricted by previous experiments and by
theoretical arguments. None of them contributes linearly at tree level to the
electroweak precision parameters. Consistency of the effective theory requires that
their size cannot significantly exceed $\left(\frac{2m_t}{\Lambda}\right)^2$, but
the new physics scale $\Lambda$ is unknown. All we know for sure is that they must be
less than unity, and this fact justifies throwing away all contributions that are
quadratic in the new couplings.

It is now a straightforward but tedious calculation to compute the contribution of
each of these operators to the differential cross section, a calculation to which we
now turn.

\section{New physics contributions to the differential cross sections}
\label{sec:diffsigma}

In this section I derive the most important result of this paper: the contribution
of each of the operators in~(\ref{eq:gtt},\ref{eq:tWb},\ref{eq:4F}) to the fully
differential cross section comprising $t\overline t$~pair production and decay.

\begin{figure}
\begin{center}
\mbox{\epsfxsize=3in \epsfbox{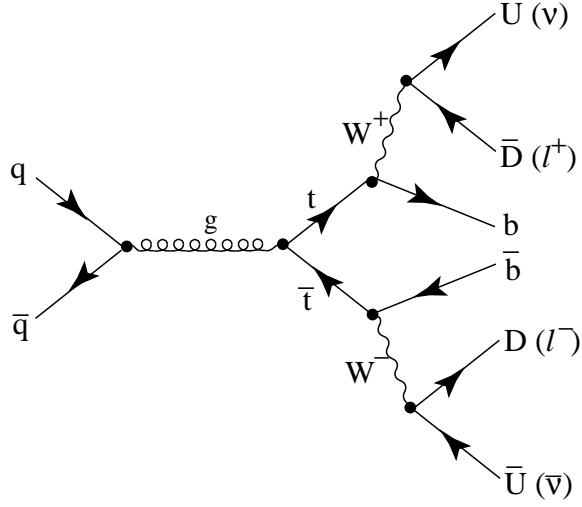}}
\end{center}
\caption{\protect\label{fig:6part}The full cascade of  $t\overline t$ production and
decay in the Standard Model. The decay products of the $W^+$ are an up-type $U=u,c$
quark  (or a neutrino) and a down-type antiquark $\overline D=\overline d, \overline
s$ (or a positively  charged lepton, $e^+$ or $\mu^+$).}
\end{figure}

The SM process is shown in~Fig.~\ref{fig:6part}. By~$D$ I designate the down-type
quark in the decay. In (semi)leptonic decays it should be replaced by a charged
lepton. In the following I will keep, for the economy of the notation,
$p_D$~for the lepton momentum; I well proceed similarly with $\overline D,U$
and~$\overline U$.

A Breit-Wigner formula is applicable to the amplitudes when the $t$ and $\overline
t$, as well as the~$W^{\pm}$ are on shell:
\begin{equation}
{\cal M}_{q\overline{q}\rightarrow 6}=
 \sum_{s,\overline s}{\cal M}_{t\rightarrow3}(s)\left(\frac{\pi\,\delta
    \left(p_t^2-m_t^2\right)}{m_t\Gamma_t}\right)^{\frac{1}{2}}
 {\cal M}_{q\overline{q}\rightarrow t\overline{t}}
  \left(\frac{\pi\,\delta\left(p_{\overline{t}}^2-m_t^2\right)}
   {m_t\Gamma_t}\right)^{\frac{1}{2}}
 {\cal M}_{\overline t\rightarrow 3}({\overline s}).
\end{equation}

This equation allows us to separate the physics into top production, described by
the density matrix $\rho_{s\overline s,s^\prime\overline s^\prime}$ and the decay
represented by $T_{ss^\prime}^{(t)}$ and $T_{\overline s\overline
s^\prime}^{(\overline t)}$:
\begin{equation}
\frac{g_s^4}{9}\times\rho_{s\overline s,s^\prime\overline s^\prime}=
\overline{\sum}| {\cal M}_{q\overline{q}\rightarrow t\overline{t}}|^2,
\end{equation}
averaged for the incoming spin and color, summed for outgoing color (but the
$t,\overline t$ spins are kept fixed).

The decay of the $t,\overline t$ quarks is described by 
\begin{eqnarray}
\sum\overline{{\cal M}_{t\rightarrow 3}\left(s^\prime\right)}
               {\cal M}_{t\rightarrow 3}\left(s\right)&=&
\frac{T_{ss^\prime}^{(t)}}{|p_W^2-m_W^2|^2}
\Longrightarrow\left(\frac{\pi\,\delta\left(p_W^2-m_W^2\right)}
   {m_W\Gamma_W}\right)\times T_{ss^\prime}^{(t)}
\nonumber\\
\sum\overline{{\cal M}_{\overline t\rightarrow 3}\left(\overline s^\prime\right)}
               {\cal M}_{\overline t\rightarrow 3}\left(\overline s\right)&=&
\frac{T_{\overline s\overline s^\prime}^{(\overline t)}}{|p_{\overline W}^2-m_W^2|^2}
\Longrightarrow\left(\frac{\pi\,\delta\left(p_{\overline W}^2-m_W^2\right)}
   {m_W\Gamma_W}\right)\times T_{\overline s\overline s^\prime}^{(\overline t)}.
\end{eqnarray}

In the decay contribution I separate out factors that do not receive contributions
from new physics operators,
\begin{eqnarray}
T_{ss^\prime}^{(t)}&=&
\left(\frac{g}{\sqrt 2}\right)^4|V_{U\overline D}|^2
\times 8\, (p_b\cdot p_U) (p_t\cdot p_{\overline
D})\times\tau_{ss^\prime}^{(t)}\nonumber\\
T_{\overline s\overline
s^\prime}^{(\overline t)}&=&
\left(\frac{g}{\sqrt 2}\right)^4|V_{\overline U D}|^2
\times 8\, (p_{\overline b}\cdot p_{\overline U}) (p_{\overline t}\cdot
p_D)\times\tau_{\overline s\overline s^\prime}^{(\overline t)}.
\end{eqnarray}

The six particle phase space becomes manageable when we describe the momenta of the
three $t$ decay products in the $t$ frame and the three $\overline t$ decay products
in the $\overline t$ frame. Denoting the three-direction of the
$W(\overline W)$ by
$\bf w( \overline w)$ and the three-direction of the $\overline D(D)$ by
$\bf n( \overline n)$in these frames, we find for the differential cross section
\begin{eqnarray}\label{eq:dsigma}\hspace{-.5in}
d\sigma&=&
\frac{\alpha_s^2\,\beta^2}{9\times 2^{24}\pi^9}\,
\frac{(m_t^2-m_W^2)^2}{m_Z^4m_W^2\Gamma_t^2\Gamma_W^2(m_t^2+m_W^2)^4}
\times\left[4\left(\frac{g}{\sqrt2}\right)^4\times
2\frac{m_t^2m_W^2}{m_t^2+m_W^2}\right]^2
\times|V_{U\overline D}|^2|V_{\overline UD}|^2
\\&&
\times d\Omega_t^{\mbox{\scriptsize (ZMF)}}
\times d\Omega({\bf n})
 \times\frac{d\Omega({\bf n})}{(1-\beta_W{\bf n\cdot w})^3}\,(p_b\cdot p_U)
\times d\Omega({\bf n})
 \times\frac{d\Omega({\bf\overline n})}{(1-\beta_W{\bf\overline n\cdot
\overline w})^3}\,(p_{\overline b}\cdot p_{\overline U})
\times Z_t
\nonumber
\end{eqnarray}
where $d\Omega_t^{\mbox{\scriptsize (ZMF)}}$ is the solid angle measure of the top
direction in the zero momentum frame (i.e. the center of mass frame of the
$t\overline t$ system), $\beta_W=\frac{m_t^2-m_W^2}{m_t^2+m_W^2}$ is the speed of
the $W$ in the $t$ frame, and the new physics corrections are all factorized into
\begin{equation}\label{eq:Zt}
Z_t=\sum_{s\overline s,s^\prime\overline
s^\prime}\frac{\rho_{s\overline s,s^\prime\overline
s^\prime}}{4}\times\tau_{ss^\prime}^{(t)}\times\tau_{\overline s\overline
s^\prime}^{(\overline t)}.
\end{equation}

The spin states corresponding to the indices $s,\overline s$ require careful
definition: because we are also using the off-diagonal elements of the density
matrix, the relative phase of the spin states must be taken into account. We achieve
this using the following conventions. The spin vectors $s,\overline s$ lay in the
scattering plane and their spatial components are back-to back to each other in
the ZMF frame. A vector $\xi^\mu$ is chosen normal to the
scattering plane, and the for a spin basis I use the so-called "off-diagonal" basis
of Ref.~\cite{MahlonParke} in which the SM contribution to the density
matrix is simple. Our conventions are best understood in the ZMF where the various
vectors are
\begin{eqnarray}
\begin{array}{ll}
p_q^\mu=\frac{\sqrt{\hat s}}{2}(1|1|0|0)^\mu&
p_{\overline q}^\mu=\frac{\sqrt{\hat s}}{2}(1|-1|0|0)^\mu\\
p_t^\mu=\frac{\sqrt{\hat s}}{2}(1|\beta\cos\Theta|\beta\sin\Theta|0)^\mu&
p_{\overline t}^\mu=\frac{\sqrt{\hat
s}}{2}(1|-\beta\cos\Theta|-\beta\sin\Theta|0)^\mu\\
\eta^\mu=\frac{-1}{\sqrt{1-\beta^2\sin^2\Theta}}
 (\beta\sin\Theta|0|1|0)^\mu\hspace{3em}\vspace{.3ex}&
\overline\eta^\mu=\frac{-1}{\sqrt{1-\beta^2\sin^2\Theta}}
 (\beta\sin\Theta|0|-1|0)^\mu
\end{array}\nonumber\\
\begin{array}{ll}
s^\mu=\frac{\gamma}{\sqrt{1-\beta^2\sin^2\Theta}}
 (\beta\cos\Theta|1-\beta^2\sin^2\Theta|\beta^2\sin\Theta\cos\Theta|0)^\mu&\\
\overline s^\mu=\frac{\gamma}{\sqrt{1-\beta^2\sin^2\Theta}}
(\beta\cos\Theta|-(1-\beta^2\sin^2\Theta)|-\beta^2\sin\Theta\cos\Theta|0)^\mu
\vspace{.3ex}\\
\xi^\mu=(0|0|0|1)^\mu&
\end{array}
\end{eqnarray}
Here, $\beta$ is the speed of the top quark in the ZMF frame,
$\gamma=1/\sqrt{1-\beta^2}$ and $\Theta$ is the top scattering angle.

With these definitions the phases of the spin states are fixed by requiring
\begin{equation}
\begin{array}{lcr}
u_{\uparrow}\overline u_{\downarrow}=+\frac{1}{2}(\gamma.p_t+m_t)\ \gamma^5\
\gamma.(\xi+i\eta)&\ \ \ \ &
u_{\downarrow}\overline u_{\uparrow}=+\frac{1}{2}(\gamma.p_t+m_t)\ \gamma^5\
\gamma.(\xi-i\eta)\\
v_{\uparrow}\overline v_{\downarrow}=-\frac{1}{2}(\gamma.p_{\overline t}-m_t)\
\gamma^5\ \gamma.(\xi-i\overline\eta)&&
v_{\downarrow}\overline v_{\uparrow}=-\frac{1}{2}(\gamma.p_{\overline t}-m_t)\
\gamma^5\ \gamma.(\xi+i\overline\eta)
\end{array}
\end{equation}
where the $\uparrow\,(\downarrow)$ states now have spin projection $+1/2\,(-1/2)$ on
the corresponding spin vector.

In the $t(\overline t)$ frame the purely spatial vectors
$\bf{\mbox{\boldmath$\eta$}},s,{\mbox{\boldmath$\xi$}}\,(\bf
\overline{\mbox{\boldmath$\eta$}},\overline s,{\mbox{\boldmath$\xi$}})$ form
right handed orthonormal bases, so that they are used as the respective $x,y,z$ axes
to help define the directions of the momenta of the decay products. Although the
differential cross section in the six-particle phase space is independent of the
choice of the $t,\overline t$ spin basis, the choice of the variables does depend on
these Cartesian frames. 

Given these conventions we can now calculate the $\rho$ and $\tau$ matrices in the
SM,
\begin{equation}\label{eq:rhoSM}
\rho_{\uparrow\downarrow,\uparrow\downarrow}^{SM}=
\rho_{\downarrow\uparrow,\downarrow\uparrow}^{SM}=2-\beta^2\sin^2\Theta
,\hspace{3em}
\rho_{\uparrow\downarrow,\downarrow\uparrow}^{SM}=
\rho_{\downarrow\uparrow,\uparrow\downarrow}^{SM}=\beta^2\sin^2\Theta,
\end{equation}
and all other elements vanish~\cite{MahlonParke}. The decay contributions are,
in the notation \(
\left[\begin{array}{cc}
\uparrow\uparrow&\uparrow\downarrow\\
\downarrow\uparrow&\downarrow\downarrow
\end{array}\right]\),
\begin{eqnarray}\label{eq:tauSM}
\tau^{(t)}_{SM}&=&
\left[\begin{array}{cc}
1+{\bf s\cdot n}&{\bf n\cdot}({\bf {\mbox{\boldmath$\xi$}}}+
 i{\bf{\mbox{\boldmath$\eta$}}})\\
{\bf n\cdot}({\bf {\mbox{\boldmath$\xi$}}}-i{\bf{\mbox{\boldmath$\eta$}}})&
 1-{\bf s\cdot n}
\end{array}\right]=
\left[\begin{array}{cc}
1+\cos\vartheta&\sin\vartheta e^{i\varphi}\\
\sin\vartheta e^{-i\varphi}&1-\cos\vartheta
\end{array}\right],
\nonumber\\
 \tau^{(\overline t)}_{SM}&=&
\left[\begin{array}{cc}
1-{\bf\overline s\cdot\overline n}&{\bf\overline n\cdot}({\bf
{\mbox{\boldmath$\xi$}}}+i{\bf\overline{\mbox{\boldmath$\eta$}}})\\ 
{\bf\overline n\cdot}({\bf
{\mbox{\boldmath$\xi$}}}-i{\bf\overline{\mbox{\boldmath$\eta$}}})&
1+{\bf\overline s\cdot\overline n}
\end{array}\right]=
\left[\begin{array}{cc}
1-\cos\overline\vartheta&\sin\overline\vartheta e^{i\overline\varphi}\\
\sin\overline\vartheta e^{-i\overline\varphi}&1+\cos\overline\vartheta
\end{array}\right].
\end{eqnarray}
Substituting~(\ref{eq:rhoSM}) and~(\ref{eq:tauSM}) into~(\ref{eq:Zt}) we find
\begin{equation}
Z_t^{(SM)}=(\,1+\cos\vartheta\cos\overline\vartheta\,)
 (1-\frac{1}{2}\,\beta^2\sin^2\Theta)
 +\frac{1}{2}\,\beta^2\sin^2\theta\sin\vartheta\sin\overline\vartheta
   \cos(\varphi-\overline\varphi).
\end{equation}

In these expressions I introduced the polar angles in the "off-diagonal" basis.
These are defined as\footnote{There is a sign difference in the definition of the
angles $\varphi,\overline\varphi$ that I am using here and the one used
in~\cite{HoldomTorma}. The difference shows up in the form of the  contribution due
to~$\kappa$.}
\begin{equation}
\begin{array}{lclcl}
{\bf n\cdot s}=\cos\vartheta &\ \ \ \ &
  {\bf n\cdot{\mbox{\boldmath$\xi$}}}=\cos\varphi\sin\vartheta &\ \ \ \ &
  {\bf n\cdot{\mbox{\boldmath$\eta$}}}=\sin\varphi\sin\vartheta\\
{\bf\overline n\cdot\overline s}=\cos\overline\vartheta &\ \ \ \ &
  {\bf\overline n\cdot{\mbox{\boldmath$\xi$}}}=
  \cos\overline\varphi\sin\overline\vartheta &\ \ \ \ &
  {\bf\overline n\cdot\overline{\mbox{\boldmath$\eta$}}}=
  \sin\overline\varphi\sin\overline\vartheta\\
{\bf w\cdot s}=\cos\chi &\ \ \ \ &
  {\bf w\cdot{\mbox{\boldmath$\xi$}}}=\cos\phi\sin\chi &\ \ \ \ &
  {\bf w\cdot{\mbox{\boldmath$\eta$}}}=\sin\phi\sin\chi\\
{\bf\overline w\cdot\overline s}=\cos\overline\chi &\ \ \ \ &
  {\bf\overline w\cdot{\mbox{\boldmath$\xi$}}}=
  \cos\overline\phi\sin\overline\chi &\ \ \ \ &
  {\bf\overline w\cdot\overline{\mbox{\boldmath$\eta$}}}=
  \sin\overline\phi\sin\overline\chi
\end{array}
\end{equation}

Any corrections to the $gt\overline t$ vertex can only show up in $\rho$ while
new physics in top decay will affect only the $\tau$'s. At this point the astute
reader may observe that while the decay contributions affect the distributions in
all the observed angles, the corrections to top production cannot change the
distribution in ${\bf w},\overline{\bf w}$. This clear distinction is, however, lost
when experimental realities are taken into account. The jets produced by the
$U,\overline D$ quarks are indistinguishable and the required symmetrization brings
back the $\bf w$-dependence. 

Now look at the contributions from each of the twelve new physics operators. I
quote here only the result of a tedious but straightforward calculation.
The contribution of the chromoelectric and chromomagnetic moments into $Z_t$ is
\begin{eqnarray}\label{eq:Zkappa}
Z_\kappa&=&
\ \ \kappa\times\left\{
 2\,(1+\cos\vartheta\cos\overline\vartheta)+\beta^2\gamma\,\sin\Theta\cos\Theta
 \,(\sin\vartheta\cos\overline\vartheta\sin\varphi+
 \sin\overline\vartheta\cos\vartheta\sin\overline\varphi)
 \right\}\nonumber\\&&
+\tilde\kappa\times\left\{
 \beta\gamma\,\sin\Theta\,\sqrt{1-\beta^2\sin^2\Theta}\,
 (\sin\vartheta\cos\overline\vartheta\cos\varphi+
 \sin\overline\vartheta\cos\vartheta\cos\overline\varphi)
\right\},
\end{eqnarray}
and the four-quark operators contribute
\begin{eqnarray}\label{eq:ZC}
Z_C&=&
\ \ c_{VA}\times\beta\gamma\,\frac{
  \gamma\,(2-\beta^2\sin^2\Theta)\cos\Theta\,(\cos\vartheta+\cos\overline\vartheta)
  -\sin\Theta\,(\sin\vartheta\sin\varphi+\sin\overline\vartheta\sin\overline\varphi)
   }{\sqrt{1-\beta^2\sin^2\Theta}}\nonumber\\&&
+c_{AA}\times\beta\gamma\left\{
  2\gamma\cos\Theta\,(1+\cos\vartheta\cos\overline\vartheta)-
  \sin\Theta\,(\sin\vartheta\cos\overline\vartheta\sin\varphi+
  \sin\overline\vartheta\cos\vartheta\sin\overline\varphi)
  \right\}\nonumber\\&&
+c_{AV}\times2\,\gamma^2\,\sqrt{1-\beta^2\sin^2\Theta}\,
  (\cos\vartheta+\cos\overline\vartheta)
  \\&&
+c_{VV}\times2\,\gamma^2\left\{
 (1+\cos\vartheta\cos\overline\vartheta)(1-\frac{1}{2}\beta^2\sin^2\Theta)+
 \frac{1}{2}\beta^2\sin^2\Theta\,
 \sin\vartheta\sin\overline\vartheta\cos(\varphi-\overline\varphi)
\right\}.\nonumber
\end{eqnarray}

These contributions, as I explained above, do not depend on the angles
$\chi,\phi,\overline\chi,\overline\phi$. The decay corrections, however, do:
\begin{eqnarray}\label{eq:Za}
Z_a&=&
\ -\mu_1\times\left\{\rule{0pt}{3ex}
  \left[
  \sin\vartheta\sin\chi\cos(\varphi-\phi)+
  \sin\overline\vartheta\sin\overline\chi\cos(\overline\varphi-\overline\phi)
  \right.\right.\nonumber\\&&\hspace{5em}\left.\left.
  +(\cos\vartheta+\cos\overline\vartheta)(\cos\chi+\cos\overline\chi)\right]
  \times (1-\frac{1}{2}\beta^2\sin^2\Theta)
  \right.\nonumber\\&&\left.\hspace{10ex}-\hspace{1ex}
  \frac{1}{2}\beta^2\sin^2\Theta\left[
  \sin\overline\vartheta\sin\chi\cos(\overline\varphi-\phi)+
  \sin\vartheta\sin\overline\chi\cos(\varphi-\overline\phi)
  \right]
  \right\}\nonumber\\&&
+\mu_2\times\left\{\rule{0in}{3ex}\right.
  \left[
  \sin\vartheta\cos\overline\vartheta\sin\chi\sin(\varphi-\phi)+
  \cos\vartheta\sin\overline\vartheta\sin\overline\chi
  \sin(\overline\varphi-\overline\phi)
  \right](1-\frac{1}{2}\beta^2\sin^2\Theta)\nonumber\\&& \hspace{4em}
  -\frac{1}{2}\beta^2\sin^2\Theta\left[
  \sin\vartheta\sin\overline\vartheta\,(\cos\chi-\cos\overline\chi)
  \sin(\varphi-\overline\varphi) \right.\nonumber\\&&\left.\hspace{10em}+
  \sin\vartheta\cos\overline\vartheta\sin\overline\chi\sin(\varphi-\overline\phi)+
  \sin\overline\vartheta\cos\vartheta\sin\chi\sin(\overline\varphi-\phi)          
  \right]
  \left.\rule{0in}{3ex}\right\}\nonumber\\&&
+\mu_1\times\left\{
  (1+\cos\vartheta\cos\overline\vartheta)
  (1-\frac{1}{2}\beta^2\sin^2\Theta)+
  \frac{1}{2}\beta^2\sin^2\Theta
  \sin\vartheta\sin\overline\vartheta\cos(\varphi-\overline\varphi)
  \right\}
\end{eqnarray}
The contribution of the CP conserving $\mu_1$ in the last line is proportional
to the SM contribution and only rescales the CKM matrix elements by a
factor $(1+\mu_1)$. One can drop this contribution. 

The form of the contributions in~Eqns.~(\ref{eq:Zkappa},\ref{eq:ZC},\ref{eq:Za})
constitutes the main result of the present paper. They can be used in an experiment
to disentengle the contributions from the various new physics operators. Integrating
over the various angles except $\Theta$ as required by Eqns.~(\ref{eq:dsigma})
immediately convinces us that all the information on many of the operators resides in
top polarization. Namely, the integration is equivalent to replacing
\begin{equation}
Z_t\,\rightarrow\,Z_t=\left(1-\frac{1}{2}\beta^2\sin^2\Theta\right)
 \left[1+\mu_1+2\gamma^2\times c_{VV}\right]+2\kappa+2\beta\gamma^2\cos\Theta
 \times c_{AA}.
\end{equation}
The only couplings that affect the distribution of the events in the $t,\overline t$
momenta at all are $\mu_1$, $c_{VV}$ and $c_{AA}$. Out of these, $c_{AA}$ does not
contribute to the total cross section.

\section{Sensitivities}
\label{sec:sens}

The parton-level calculations in Sec.~\ref{sec:diffsigma} should be confronted with
experimental realities. At first sight one would think that the large inaccuracies
related to the extraction of parton momenta from jet observables preclude a detailed
mapping of the phase space. However, in order to establish the presence of the new
physics operators one does not need to reconstruct the full angular distribution.
The complicated structure of Eqns.~(\ref{eq:Zkappa},\ref{eq:ZC},\ref{eq:Za}) does
not allow us to select a simple observable that would indicate the presence of new
physics. Instead, one should look for the particular types of angular
correlations that are similar to those in the above equations. This is most easily
done by a Maximum Likelihood estimate. Following this strategy the angular
correlations are not spoiled by uncertainties in the jet variables even as large as
$\pm50\%$ or by very large backgrounds~\cite{HoldomTorma}.

The events fall into three categories, according to the decay mode of the
$t,\overline t$:
\begin{itemize}
\item{Dilepton events. In this case two neutrinos are produced and full event
reconstruction is extremely hard. The angles we are investigating are probably
impossible to extract.}
\item{Semileptonic events. These provide an opportunity to accurately measure the
momentum of the charged lepton (i.e. $p_{\overline D}$ or $p_{D}$}), but on the
hadronic side the two light quark jets are practically indistinguishable. One must
symmetrize the predicted differential cross sections and compare the resulting
quantity to the experiment. Fortunately, we will see that this procedure does not
entail much information loss. However, the smaller leptonic top decay branching
ratio results in smaller statistics compared to the all-hadronic decays.
\item{All hadronic events. In this case the advantages of high statistics are
somewhat compensated by the hardships of relating six jets to six partons.
Nevertheless, two SVX tags may identify the $b,\overline b$ quarks and requiring
that the both tops and W's are on shell leaves a chance for an approximate event
reconstruction.

One is faced, however, with more complicated ambiguities in this case. There are two
pairs of indistinguishable light quarks jets plus the entire $t$ side is hard to
distinguish from the $\overline t$ side. The former problem does not introduce large
uncertainties in the estimation of new physics parameters. As for the latter, we
observe that the effect of $t\,\Leftrightarrow\,\overline t$ replacement amounts to
the replacement of $\alpha\,\leftrightarrow\,\pi-\overline\alpha$ of all the angles
in the above formulas for $Z_t$. Consequently, some of the operators (namely,
$c_{AA}$ and $c_{AV}$) are antisymmetric under this replacement and cancel out from
the symmetrized cross sections. The rest however, including both operators in top
decay, are symmetric and survive unchanged.}
\end{itemize}

In order to assess the statistical uncertainties involved in the determination of
the new physics parameters I performed a Maximum Likelihood estimation of the eight
real parameters~$\mu_i=\mu_1,\mu_2,\kappa,\tilde\kappa,c_{VV},c_{VA},c_{AV},c_{AV}$.
I used a Monte Carlo generator to produce events according to the SM
distribution at $\sqrt s=2\,TeV$ and built a Maximum Likelihood estimator that finds
the most probable values  of the $\mu_i$'s (the true values are obviously
zero.)\footnote{
This analysis, an extension to many parameters of the one we used
in~\cite{HoldomTorma}, leads to the "best estimator"
\(
\frac{1}{Z_t^{SM}}\frac{\partial}{\partial\mu_i}Z_t(\mu_i).
\)
After the publication of~\cite{HoldomTorma} I realized that this is the same
estimator as the one described in~\cite{AtwoodSoni}, found through a different
argument unrelated to Maximum Likelihood.}
The fall of the probability by a factor of $e^{-1/2}$ was
interpreted as a ``$1-\sigma$" interval (in the following I will call its size two
times the ``accuracy".) In each case the ``accuracies" agreed well with the ones
found from a second-order expansion of the log-likelihood function, so that the use
of a covariance matrix $C_{ij}$ is justified:
\begin{equation}
\ln{\cal L({\bf \mu})}=\ln{\cal L}_{max}-\frac{1}{2}{\bf\mu}_iC_{ij}{\bf\mu}_j.
\end{equation}

As a check of the reliability of these estimates, I observed that in each direction
in the parameter space when only one $\mu_i$ is kept nonzero, the likelihood quickly
decreases as $\mu_i$ is moved outside the ``$1-\sigma$" interval. The true value
$\mu_i=0$ has always been consistent with the ``accuracy." In addition, using
different event numbers (all above ${\cal N}\geq1200$), I~observed that the accuracy
is indeed proportional to ${\cal N}^{-1/2}$.
\begin{center}
\begin{table}[htb]
\begin{tabular}{c||@{}c|@{}c|@{}c|@{}c||@{}c|c|c|c|c|c|c|c|c|c}
$\mu_i$&CP&$?$&?&No
&\multicolumn{2}{c|}{No Symm.}
&\multicolumn{2}{c|}{$U\leftrightarrow\overline D$}
&\multicolumn{2}{c|}{All Symm.}
&\multicolumn{2}{c|}{No Pol.}
&\multicolumn{2}{c}{No Pol., $t\leftrightarrow\overline t$}
\\\cline{6-15}&
&$\sigma_{tot}$&$t\leftrightarrow\overline t$&Pol.&
$\ \ \ \ 20k$&   $2k$&
$20k$&   $2k$&
$20k$&   $2k$&
$20k$&   $2k$&
$20k$&   $2k$
\\\hline\hline
$\mu_1$&
+&yes&yes&&
$1.4$&$4.2$&$1.5$&$\bf 4.6$&$1.7$&$5.1$&&&&\\
\tableline
$\mu_2$&
$-$&&yes&&
$3.4$&$11$&$4.4$&$\bf 14$&$5.6$&$18$&&&&\\
\hline
$\kappa$&
$+$&yes&yes&yes&
$1.7$&$5.8$&$2.0$&$\bf 6.8$&$2.1$&$7.2$&$2.8$&$27$&$2.8$&$27$\\
\hline
$\tilde\kappa$&
$-$&&yes&&
$2.0$&$6.5$&$2.6$&$\bf 8.6$&$3.3$&$11$&&&&\\
\hline
$c_{VV}$&
$+$&yes&yes&yes&
$0.4$&$1.2$&$0.4$&$\bf 1.2$&$0.4$&$1.2$&$0.4$&$1.3$&$0.4$&$1.3$\\
\hline
$c_{VA}$&
$-$&&yes&&
$0.4$&$1.1$&$0.4$&$\bf 1.2$&$0.4$&$1.3$&&&&\\
\hline
$c_{AV}$&
$-$&&&&
$0.2$&$0.8$&$0.3$&$\bf 0.9$&&&&&&\\
\hline
$c_{AA}$&
$+$&&&yes&
$0.3$&$1.0$&$0.3$&$\bf 1.0$&&&$0.4$&$1.7$&&\\
\hline\hline
$\kappa-0.09c_{VV}$&
$+$&yes&yes&yes&
$1.9$&$6.2$&$2.2$&$\bf 7.6$&$2.4$&$8.4$&$3.6$&$29$&$3.6$&$29$\\
\hline
$c_{VV}+0.09\kappa$&
$+$&yes&yes&yes&
$0.4$&$1.2$&$0.4$&$\bf 1.2$&$0.4$&$1.2$&$0.4$&$1.9$&$0.4$&$1.9$
\end{tabular}
\caption{\protect\label{table1} The statistical accuracy of the Maximum Likelihood
estimate of each parameter~$\mu_i$. The numbers mean 100 times the half-width of the
``1-sigma" interval. The $2^{nd}$, $3^{rd}$ and $4^{th}$ columns tell if the operator
contributes to the total cross section, whether it survives
$t\leftrightarrow\overline t$ symmetrization or the integration over all decay
angles except $\Theta$ (called ``No Pol."). The $2k$,~$20k$ refer to the number of
events considered. ``All Symm." denotes symmetrization for
$D\leftrightarrow\overline D$, $U\leftrightarrow\overline U$ and
$t\leftrightarrow\overline t$. The column in boldface applies directly to
semileptonic events at Tevatron Run~2.}
\end{table}\end{center}

The accuracies of the results of this ML estimate are shown in Table~\ref{table1}
for 2000 and 20,000 events. The former is relevant for the semileptonic
events at the Fermilab Tevatron, while the latter is a hypothetical number shown in
order to expose the gain in the accuracy when more events are observed. These
numbers immediately confirm that the indistinguishability of the light quark jets
does not drastically deteriorate the extraction, contrary to what one might have
na\"\i vely expected. 

The intricate pattern of correlations induced by the contributions to $Z_t$ tells us
that the much of the information on the new operators resides in the spins of
the~$t,\overline t$. When all the angles related to top decay are integrated out and
only the variables $\beta$ and $\Theta$ are kept, a reduced amount of information is
still available. However, the only operators whose contributions do not vanish in
this case are $\kappa,c_{AA}$ and $c_{VV}$.  We see in~Table~\ref{table1} that in the
case of $c_{AA}$ and $c_{VV}$, much information resides in the
$\beta,\Theta$-distributions and these two operators can be observed without
measuring the angular correlations of the decay products. In fact $c_{VV}$
represents only a momentum-dependent, ${\cal O}(p^2)$ correction to the strong
coupling of the top, so its contribution is proportional to that of the SM process
times~$\gamma^2$. This different dependence on the total energy allows that we see
this operator in the unpolarized $t\overline t$ cross section. In the case of
$\kappa$, the accuracy drastically deteriorates when we integrate out the
polarization information, in accordance with the findings in Ref.~\cite{ouracc}.

The difference in the patterns of angular correlations introduced by each operator
leads to the fact that, with one exception, the principal axes of the covariance
matrix approximately coincide with the chosen operator basis (i.e. $C_{ij}$ is
approximately diagonal.) This means that the angular correlations provide almost
independent measurements of each operator in our basis. In the case of $\kappa$ and
$c_{VV}$, there is an ${\cal O}(10\%)$ mixing and the combinations corresponding to
the principal axes are shown in the last two rows of Table~\ref{table1}.

The accuracies attainable in this way, supposing that systematic uncertainties will
not play a drastic role, are generically $\pm0.01$~to~$\pm0.1$. This can be
translated to an observable new physics scale of $\Lambda={\cal O}(6-20)m_t$,
with an uncertain numerical factor of order unity.\footnote{The normalization of the
couplings~$\mu_i$ is so chosen that all mass dimensions are removed at the scale of
$\sim2\,m_t$. Most $t\overline t$ pairs are produced with close to threshold
energies, due to the increase of the parton distribution functions at small~$x$, so
that $m_t$ is the only energy scale in the problem. However, there remains some
${\cal O}(1)$ arbitrariness in the normalization.} The 4-quark operators can be
measured with a precision of~$\pm0.01$ in semileptonic events, as far as only
statistical inaccuracies are taken into account. Two of these, $c_{AV}$ and $c_{AA}$
are theoretically disfavored in many models as they involve an axial vector current
of light quarks. The other two, $c_{VV}$ and $c_{VA}$ may however remain
unsuppressed as they are related to vector currents of the light quarks. Of course,
the actual suppression of each operator is a model-dependent question and should not
be addressed in this paper.

\section{Conclusions and prospects of discovery}

In the above we found that four CP-conserving and four CP-violating new physics
operators can contribute to top pair production in interference with 
$q\overline q\rightarrow t\overline t$: the chromomagnetic and chromoelectric
moments $\kappa,\tilde\kappa$, two decay corrections $\mu_0,\mu_1$ and the four
current-current color octet 4-quark operators
$c_{VV},c_{VA},c_{AV},c_{AA}$ (averaged over the light quark flavors.) In deriving
our set of contributing operators I used the equations of motion, so that some of
the physics in the $gt\overline t$ couplings may actually show up in the four-quark
operators (for example, $G_{\mu\nu}^a\,\overline
t\frac{\lambda^a}{2}\gamma^\mu D^\nu\,b$ is eliminated.)  All these operators arise
in both the linear and the nonlinear (chiral) set, so the two cases cannot be told
apart in this process. I calculated the contribution of each of these operators to
the fully differential cross section
$d\sigma(q\overline q\rightarrow t\overline t\rightarrow6\,fermions)$, at tree level,
in the approximation when all $t,\overline t,W^+,W^-$ are on shell. This is the
dominant process of top pair production on the Tevatron at $\sqrt s=2\,TeV$. At
higher energies (for example at the LHC) our analysis is not relevant, because other
Standard Model processes become important.

Using a maximum likelihood method I estimated the statistical inaccuracy of the
measurement of this parameter set. We find that only  $c_{AA}$ and $c_{VV}$ can be
meaningfully measured without using the information encoded in the directions of the
momenta of the six decay products. In addition, the contribution of all operators
except $c_{VV}$ to the total cross section vanishes. It is well known that the total
cross section is not measured very precisely at the Tevatron, due to inaccurate
knowledge of the luminosity. This fact can affect therefore only the extraction of
$c_{VV}$ but not that of the other parameters.

The statistical inaccuracies in the case of 2000 semileptonic events (feasible at
the Tevatron), generically~$\pm{\cal~O}(0.01-0.1)$, correspond to new physics on
the TeV scale. The actual assessment of the meaning of this accuracy should be
performed in the context of particular models, but in any case these accuracies are
an order of magnitude better than those previously found without the use of spin
information. Although these results seem to be quite robust against large
backgrounds or inaccuracies in event reconstruction, it remains to be seen how large
is the effect of asymmetric systematic distortions of the distributions by
experimental factors. This important question should be addressed by detailed
simulations of the experiments, which the present author is not equipped to perform.

As it was observed in~\cite{HoldomTorma}, when the event number is less than
$\sim1000$, the estimation becomes useless due to very large inaccuracies
exceeding the above quoted unitarity bounds. The small branching ratio of the
leptonic top decay renders dilepton events useless for our purposes. The importance
of the event number to improve the statistics leads us to consider the abundant
all-hadronic decays modes. We have shown that the price to pay is the loss of the
measurement of $c_{AV}$ and $c_{AA}$, but the increased event number can help to
improve the accuracy of the rest of the parameters.

\section*{Acknowledgments}

I would like to thank Alakhaba Datta for useful discussions. This work was
supported by the Natural Sciences and Engineering  Research Council of Canada.

\appendix
\section{Linear representation}\label{app:lin}

In the linear representation there are no dimension~4 or~5 operators. In the
following I provide a list of those operators with two quark fields, at dimension
six, which in the unitary gauge generate vertices involving at least one top quark
(terms proportional with light or $b$ quark masses, and also all lepton fields, have
been dropped):

\begin{eqnarray}
b_1^{\alpha\beta}&\times&(\overline q^\alpha_L\,i D_\mu
     d^\beta_R)\,iD^\mu\Phi+h.c.\\ 
b_2^{\alpha\beta}&\times&(\overline q^\alpha_L\,i D_\mu
     u^\beta_R)\,\epsilon\,\overline{iD^\mu\Phi}+h.c.\\ 
b_3^{\alpha\beta}&\times&(\overline q^\alpha_L\,\gamma_\mu
     q^\beta_L)\,(\overline\Phi\,iD^\mu\Phi)+h.c.\\ 
b_4^{\alpha\beta}&\times&(\overline q^\alpha_L\,\gamma_\mu\frac{\tau^a}{2}
     q^\beta_L)\,(\overline\Phi\,\frac{\tau^a}{2}\,iD^\mu\Phi)+h.c.\\
b_5^{\alpha\beta}&\times&(\overline u^\alpha_R\,\gamma_\mu
     u^\beta_R)\,(\overline\Phi\,iD^\mu\Phi)+h.c.\\ 
b_6^{\alpha\beta}&\times&(\overline u^\alpha_R\,\gamma_\mu
     d^\beta_R)\,(\Phi^T\epsilon\,iD^\mu\Phi)+h.c.\\ 
b_7^{\alpha\beta}&\times&(\overline q^\alpha_L\,
     u^\beta_R\,\epsilon\overline\Phi^T)\,(\overline\Phi\Phi)+h.c.\\ 
g^\prime\,b_8^{\alpha\beta}&\times&\overline q^\alpha_L\,
     \sigma^{\mu\nu}B_{\mu\nu}u^\beta_R\ \epsilon\overline\Phi^T+h.c.\\ 
g\,b_9^{\alpha\beta}&\times&\overline q^\alpha_L\,\sigma^{\mu\nu}u^\beta_R
     \ \frac{\tau^a}{2}V^a_{\mu\nu}\ \epsilon\overline\Phi^T+h.c.\\
g_s\,b_{10}^{\alpha\beta}&\times&\overline q^\alpha_L\,\sigma^{\mu\nu}
     \ \frac{\lambda^a}{2}G^a_{\mu\nu}\ u^\beta_R\ \epsilon\overline\Phi^T+h.c.\\ 
g\,b_{11}^{\alpha\beta}&\times&\overline q^\alpha_L\,\sigma^{\mu\nu}
     d^\beta_R\ \frac{\tau^a}{2}V^a_{\mu\nu}\ \Phi+h.c.
\end{eqnarray}

Here, $\alpha,\beta$ are flavor indices, $q_L^\alpha$ are the left handed quark
doublets, $u_R,d_R$ are the right-handed singlet quarks, $B_{\mu\nu}$ and
$V_{\mu\nu}$ are the hypercharge and $SU(2)$ field strengths, $G^a_{\mu\nu}$ is the
gluon field strength, and $\epsilon$ is the antisymmetric $2\times2$ tensor in
$SU(2)$ space. The covariant derivative $D_\mu$ includes all electroweak and
gluon fields.  The coupling constants, $b_i$, are (not necessarily Hermitian)
complex matrices. In the above expressions the equations of motion have been used.
This is consistent if one keeps the 4-quark operators~\cite{Whisnant}. In the unitary
gauge, replacing $\Phi\rightarrow(0,\frac{v+H}{\sqrt2})$ allows us to write down the
new physics operators in a physical basis where Goldstone fields are absent. 

There is a restriction on the size of $b_7$ from the fact that it gives mass to the
quarks. The absence of fine tuning requires that this additional mass,
$\frac{b_7^{\alpha\beta}}{2}\left(\frac{v}{\sqrt2}\right)^3$, does not significantly
exceed the physical quark masses. This coupling, however, will not give a
contribution to our process.
The coupling $b_4$ contains a piece 
\(
-\frac{b_4^{\alpha\beta}}{2}\,\frac{g}{\sqrt2}\,\,\overline u_L^\alpha \gamma^\mu
W_\mu d_L^\beta
\)
whose effect is to modify the corresponding CKM matrix element $V_{\alpha\beta}$.
Again, the absence of fine tuning imposes a restriction on the size of $b_4$.
There are obvious restrictions of the flavor changing operators in the list.

\section{Nonlinear representation}\label{app:nonlin}

In the nonlinear representation we have the following operators with two quark
fields, again dropping those that contain no top quark:
\begin{itemize}
\item{Dimension~4}
\begin{eqnarray}
&&-\frac{g}{2 \cos\theta}\left[
\lambda_{ZLu}^{\alpha\beta} \times Z_\mu\,\overline u^\alpha_L\gamma^\mu u^\beta_L+
\lambda_{ZLd}^{\alpha\beta} \times Z_\mu\,\overline d^\alpha_L\gamma^\mu d^\beta_L+
\lambda_{ZRu}^{\alpha\beta} \times Z_\mu\,\overline u^\alpha_R\gamma^\mu u^\beta_R
\right.\nonumber\\&&\left.\hspace{5em}+
\lambda_{ZRd}^{\alpha\beta} \times Z_\mu\,\overline d^\alpha_R\gamma^\mu d^\beta_R
\right]
-\frac{g}{\sqrt2}\left[\lambda_{WR}^{\alpha\beta} \times W_\mu\,\overline
  u^\alpha_R\gamma^\mu d^\beta_R+h.c.\right]
\end{eqnarray}
and $\lambda_{WL}^{\alpha\beta}$ is absorbed into the CKM matrix
element $V_{\alpha\beta}$. Here, each $\lambda_{Z}$ is Hermitian but $\lambda_{WR}$
can be any complex matrix.
\item{Dimension 5}\\
The magnetic moments are
\begin{equation}
-\overline u^\alpha\sigma^{\mu\nu}\frac
{
\kappa_\gamma^{\alpha\beta}eF_{\mu\nu}
+\kappa_g^{\alpha\beta}g_s\frac{\lambda^a}{2}G^a_{\mu\nu}
+\kappa_Z^{\alpha\beta}\frac{g}{2\cos\theta}Z_{\mu\nu}
}
{4m_t}u^\beta
-\left\{\overline u^\alpha\sigma^{\mu\nu}d^\beta\frac{\kappa_W^{\alpha\beta}
\frac{g}{\sqrt2}W_{\mu\nu}}{4m_t} +h.c.\right\},
\end{equation}
the electric moments are found from these with the replacement
$\kappa\rightarrow\tilde\kappa,
X_{\mu\nu}\rightarrow{\rule{0pt}{1ex}^*}X_{\mu\nu}$.\\
The weak derivative couplings are
\begin{equation}\hspace{-1pt}
-\frac{g}{2\cos\theta}Z^\mu\,\frac{\kappa_Z^{\prime\alpha\beta}}{2m_t}\overline
u^\alpha_L\,i\!\stackrel{\leftrightarrow}{D}_\mu u_R^\beta
 -\left\{
\frac{g}{\sqrt2}W^\mu\left(
\frac{\kappa^{\prime\alpha\beta}_{RW}}{4m_t}\ \overline
u^\alpha_L\,i\stackrel{\leftrightarrow}{D}_\mu d^\beta_R 
+\frac{\kappa^{\prime\alpha\beta}_{LW}}{4m_t}\ \overline
u^\alpha_R\,i\stackrel{\leftrightarrow}{D}_\mu d^\beta_L
\right)+h.c.\right\}.
\end{equation}
Again,
$\kappa_\gamma,\kappa_g,\kappa_Z,\tilde\kappa_\gamma,\tilde\kappa_g,\tilde\kappa_Z$
are Hermitian and $\kappa^\prime_Z,\kappa_{LW}^\prime,\kappa_{RW}^\prime$ need not
be. Because $\kappa_W$ and $\tilde\kappa_W$ are not independent,
$\tilde\kappa_W{\rule{0pt}{1.5ex}^*W_{\mu\nu}\overline
u_R\sigma^{\mu\nu}d_L\equiv i\tilde\kappa_WW_{\mu\nu}\overline
u_R\sigma^{\mu\nu}d_L}$, we can require that $\kappa_W$ and $\tilde\kappa_W$ be both
Hermitian.
\item{dimension~6}\\
Dropping all operators that involve, in addition to two quark fields, at least two
more electroweak bosons, we find for both $H=L,R$:
\begin{eqnarray}
\frac{g}{2\cos\theta}\,\frac{\rho^{\alpha\beta}_{HZ}}{m_t^2}&\times&
  \overline u_H^\alpha \gamma^\mu\,i\stackrel{\leftrightarrow}{D}^\nu
   u_H^\beta\left( \partial_\mu Z_\nu+\partial_\nu Z_\mu\right)\\
\frac{g}{\sqrt2}\,\frac{\rho^{\alpha\beta}_{HW}}{m_t^2}&\times&
  \overline u_H^\alpha \gamma^\mu\,i\stackrel{\leftrightarrow}{D}^\nu
   d_H^\beta\left( \partial_\mu W_\nu+\partial_\nu W_\mu\right)\\
\frac{g\,g_s}{2\cos\theta}\,\frac{\rho_{HZ}^{\prime\,\alpha\beta}}{m_t^2}&\times&
  \,Z_\mu\, \overline u_H^\alpha\, \frac{\lambda^a}{2}G^{\mu\nu}_a\, \gamma_\nu 
   u_H^\beta\\
\frac{g\,g_s}{\sqrt2}\,\frac{\rho_{HW}^{\prime\,\alpha\beta}}{m_t^2}&\times&
  \,W_\mu\, \overline u_H^\alpha\, \frac{\lambda^a}{2}G^{\mu\nu}_a\, \gamma_\nu 
   d_H^\beta\,+h.c.\\
\frac{g\,g_s}{2\cos\theta}\,\frac{\tilde\rho_{HZ}^{\prime\,\alpha\beta}}{m_t^2}&\times&\,Z_\mu\,
   \overline u_H^\alpha\,\frac{\lambda^a}{2}{\rule{0pt}{1.5ex}^*}G^{\mu\nu}_a\,
   \gamma_\nu u_H^\beta\\
\frac{g\,g_s}{\sqrt2}\,\frac{\tilde\rho_{HW}^{\prime\,\alpha\beta}}{m_t^2}&\times&
  \,W_\mu\, \overline u_H^\alpha
  \,\frac{\lambda^a}{2}{\rule{0pt}{1.5ex}^*}G^{\mu\nu}_a
  \, \gamma_\nu d_H^\beta\,+h.c.
\end{eqnarray}
Here, $\rho_{HZ},\rho_{HZ}^\prime,\tilde\rho_{HZ}^\prime$ are Hermitian but
$\rho_{HW},\rho_{HW}^\prime,\tilde\rho_{HW}^\prime$ need not be.
\end{itemize}

Note that in the nonlinear representation the covariant derivative $D_\mu$ involves
only the gluon and electromagnetic fields but not the $W$ and $Z$. All combinations
with these fields should be included as there is no restriction from EW gauge
invariance.


\thebibliography{99}
\bibitem{Rizzo}{D. Atwood, A. Kagan and T.G.~Rizzo,  Phys. Rev. {\bf D52}, 6264
(1995), hep-ph/9407408}
\bibitem{Hikasa}{K.-I. Hikasa, K. Whisnant, J.M.~Yang and B.-L.~Young, Phys.Rev.
{\bf D58}, 114003 (1998), hep-ph/9806401.}
\bibitem{HillParke} C.T.~Hill and S.J.~Parke, Phys. Rev. {\bf D49}, 4454 (1994).
\bibitem{Peccei2}R.D.~Peccei, S.~Peris and X.~Zhang, Nucl. Phys. {\bf B349}, 305
(1991).
\bibitem{Cheung} K.~Cheung, Phys. Rev. {\bf D53}, 3601 (1996), hep-ph/9610368
\bibitem{HoldomTorma}{B. Holdom, T. Torma, Phys.Rev. {\bf D60}, 114010 (1999),
hep-ph/9906208}
\bibitem{Peccei1} Peccei and X. Zhang, Nucl. Phys. {\bf B337}, 269 (1990).
\bibitem{Burgess} C.P.~Burgess and D.~London,  Phys.Rev. {\bf D48}, 4337 (1993),
hep-ph/9203216.
\bibitem{Whisnant}{K.~Whisnant, J.M.~Yang, B.-L.~Young and X.~Zhang, Phys.Rev.
{\bf D56}, 467 (1997), hep-ph/9702305.}
\bibitem{more lists}{
C.J.C.~Burges and H.J.~Schnitzer, Nucl. Phys. {\bf B228}, 454 (1983);
C.N.~Leung, S.T.~Love and S.~Rao, Z. Phys. {\bf C31}, 433 (1986);
W. Buchm\"uller and D.~Wyler, nucl. Phys. {\bf B268}, 621 (1986);
G.J.~Gounaris, F.M.~Renard and C.~Verzegnassi, Phys. Rev. {\bf D52}, 451 (1995);
G.J.~Gounaris, D.T.~Papadamou and F.M.~Renard, hep-ph/9609437;
\mbox{B.-L.~Young}, {\it Beijing, Heavy Flavor 1995}, p.~286, hep-ph/9511282.}
\bibitem{MahlonParke}{G.~Mahlon and S.~Parke, Phys. Lett. {\bf B441}, 173
(1997), hep-ph/9706304.}
\bibitem{AtwoodSoni} D.~Atwood and S.~Soni, Phys. Rev. {\bf D45}, 2405 (1991).
\bibitem{ouracc}{See~\cite{HoldomTorma}, see also~\cite{Rizzo} for the deterioration
of the accuracy.}
\end{document}